\pdfoutput=1

\documentclass[journal]{new-aiaa}
\usepackage{bm}
\usepackage{subfigure}
\usepackage{graphicx}
\usepackage{amsmath}
\usepackage{siunitx}
\usepackage{algorithm}
\usepackage{algorithmicx}
\usepackage{algpseudocode}

\title{Conservative interpolation of aeroacoustic sources in a hybrid workflow applied to fan}

 \author{
  Stefan J. Schoder\footnote{Postdoc, Institute of Mechanics and Mechatronics, Vienna. E-Mail: stefan.schoder@tuwien.ac.at}, %
  Clemens Junger\footnote{Doctoral candidate, Institute of Mechanics and Mechatronics, Vienna.}, %
  Michael Weitz\footnote{Master student, Institute of Mechanics and Mechatronics, Vienna.}, %
 \ and Manfred Kaltenbacher\footnote{Full Professor, Institute of Mechanics and Mechatronics, Vienna, and AIAA Member Senior.}}
 \affil{Vienna University of Technology, Vienna, Austria, 1060}

\usepackage{booktabs}
\usepackage[]{units}
\usepackage{multicol}

\usepackage{xcolor}

\begin{document}

\maketitle

\begin{abstract}
In low Mach number aeroacoustics, the well known disparity of scales makes it possible to apply efficient hybrid simulation models using different meshes for flow and acoustics, which leads to a powerful computational procedure. Our study applies the hybrid workflow to the computationally efficient perturbed convective wave equation with only one scalar unknown, the acoustic velocity potential. The workflow of this aeroacoustic approach is based on three steps: 1. perform unsteady incompressible flow computations on a sub-domain; 2. compute the acoustic sources; 3. simulate the acoustic field using a mesh specifically suited for computational aeroacoustics. In general, hybrid aeroacoustic methods seek for robust and conservative mesh-to-mesh transformation of the aeroacoustic sources while high computational efficiency is ensured. In this paper, we investigate the accuracy of a cell-centroid based conservative interpolation scheme compared to the more accurate cut-volume cell approach and their application to the computation of rotating systems, namely an axial fan. The capability and robustness of the cut-volume cell interpolation in a hybrid workflow on different meshes are investigated by a grid convergence study. The results of the acoustic simulation are in good agreement with measurements thus demonstrating the applicability of the conservative cut-volume cell interpolation to rotating systems.
\end{abstract}

\section*{Nomenclature}

\begin{tabbing}
  XXXX \= XX \= \kill
  $\hat{D}_{\rm RHS}$ \> = \> Amplitudes of DFT \\
  $E^\textrm{a}$ \> = \> Cell of the acoustic grid\\
  $E^\textrm{f}$ \> = \> Cell of the flow grid\\
  $F^\textrm{f}$ \> = \> Loads on the flow grid \\
  $\textrm{Ma}$ \> = \> Mach number\\
  $M^\textrm{f}$ \> = \> Number of flow cells\\
  $N^\mathrm{a}$ \> = \> FE basis function on the acoustic grid\\
  $N^\textrm{f}$ \> = \> FE basis function on the flow grid\\
  $N_\textrm{CAA}$ \> = \> Number of nodes on edges of CAA mesh\\
  $N_\textrm{CFD}$ \> = \> Number of nodes on edges of CFD mesh\\
  $\textrm{RHS}$ \> = \> FE nodal right hand side values \\
  $V_\textrm{c}$ \> = \> Volume of the intersection polyhedron, $\mathrm{m}^3$\\
  $X_q$ \> = \> Local scattered data patch\\
  $c$ \> = \> Speed of sound, m/s\\
  $e_\textrm{r}$ \> = \> Energy ratio \\
  $f^\mathrm{a}$ \> = \> Aeroacoustic source density on acoustic grid\\
  $f^\textrm{f}$ \> = \> Aeroacoustic source density on flow grid\\
  $l$ \> = \> Characteristic length of a vortex, m\\
  $p$ \> = \> Pressure, Pa\\
  $\overline{p}$ \> = \> Mean pressure, Pa\\
  $p^\mathrm{a}$ \> = \> Acoustic part of the pressure, Pa\\
  $p^\textrm{ic}$ \> = \> Incompressible part of the pressure, Pa\\
  $t$ \> = \> Time, s\\
  $\bm v$ \> = \> Velocity, m/s\\
  $\overline{\bm v}$ \> = \> Mean velocity, m/s\\
  $\bm v^\mathrm{a}$ \> = \> Particle velocity, m/s\\
  $\bm v^\textrm{ic}$ \> = \> Incompressible part of the velocity, m/s\\
  $\bm v_\textrm{r}$ \> = \> Relative velocity of the grid, m/s\\
  $\bm x$ \> = \> Global position, source position\\
  $\bm x_\textrm{c}$ \> = \> Volumetric center of intersection polyhedron\\
  $\bm z$ \> = \> Evaluation position\\
  $x,y,z$ \> = \> Spatial coordinates, m\\
  $\Gamma$ \> = \> Mesh ratio\\
  $\varepsilon$ \> = \> Coverage ratio \\
  $\lambda$ \> = \> Wave length, m\\
  $\bm \xi$ \> = \> Local coordinate\\
  $\rho$ \> = \> Density, $\mathrm{kg/m}^3$\\
  $\overline{\rho}$ \> = \> Mean density, $\mathrm{kg/m}^3$\\
  $\rho^\mathrm{a}$ \> = \> Acoustic part of the density, $\mathrm{kg/m}^3$\\
  $\rho_1$ \> = \> Density correction, $\mathrm{kg/m}^3$\\
  $\psi^\mathrm{a}$ \> = \> Acoustic velocity potential, $\mathrm{m}^2/\mathrm{s}$\\
 \end{tabbing}

\section{Introduction}
\lettrine{A} key challenge in computational aeroacoustics (CAA) is the huge
disparity of scales between flow structures and audible acoustic
wavelengths in low Mach number applications. A direct numerical simulation based on the compressible
flow equations, resolving both scales from the sound generating vortices to
the desired evaluation position, not only leads to a very high number
of cells but also to a very small time step size to minimize
dissipation of the acoustic waves. Hybrid schemes in CAA separate the
flow from the acoustic computation using aeroacoustic analogies
(see, e.g., \cite{Crighton:92,Wagner2007}). 
Thereby, an optimal computational grid can be used for each individual physical field. As a result, the two grids may
be quite different according to the following criteria: 1. near
walls, the flow grid needs refinement to resolve boundary layers;
2. the flow grid is mostly coarsened towards outflow
boundaries to dissipate vortices; 3. the acoustic grid has to
transport waves and therefore needs a uniform grid size 
all over the computational domain. To treat these different meshes in the hybrid workflow, the fundamental
requirement is an accurate data transfer from the flow to the acoustic
grid in order to minimize interpolation errors. To cope with this
task, different strategies can be applied, starting from low
complexity nearest neighbor interpolation to complex volume
intersections between flow and acoustic grid. Hybrid CAA schemes,
which consider just a forward coupling of the flow to the acoustic
field, are only valid for low Mach number flows, where the scaling
between the acoustic wavelength $\lambda$ and the characteristic length
of a vortex $l$ is given by
\[
\lambda \sim \frac{l}{\textrm{Ma}}\,,
\]
where $\textrm{Ma}$ is the Mach number. 
Therefore, the sizes of the two grids are
in general quite different and a simple nearest neighbor interpolation for
computing the acoustic sources fails (see, e.g.,
\cite{Caro2009,Schroeder2016}). In \cite{Caro2009}, the acoustic right hand side is 
computed by summing the contributions of all flow cells belonging to
finite elements surrounding an acoustic finite element node. Thereby,
it is assumed that the flow quantities used for the acoustic source
term computation are constant over each flow cell. This approach has
also been used in \cite{Piellard09} for three-dimensional problems,
where a grid dependency has been observed resulting in a too low sound
pressure level over the whole frequency spectrum. 
Any function based interpolation technique to construct a continuous function space (e.g. radial basis function interpolation \cite{hardy1971multiquadric,hardy1990theory,rendall2008unified,cordero2014radial}, finite element interpolation \cite{kaltenbacher10:2})
will improve the interpolation result while increasing the computational effort.
Particularly important for finite element method (FEM) interpolation is that
an interpolation of the sources of the equation can be done in two ways.
First, the field variable is interpolated to the integration points of the finite element
and then processed by the finite element integration. 
Second, the whole source integral is interpolated and therefore the integral 
properties are conserved leading to a natural conservative interpolation (addressed in Sec.
\ref{Sec:Formulation})
that has the same computational burden as a field interpolation.
The conservative approaches presented herein follow the idea that the integral 
properties are conserved; however, we assume that a lowest order function space for the flow field is sufficient, which leads to a fast and efficient algorithm.
If a higher order representation of the values inside the flow cells is desired, an additional interpolation and regression step must be executed to construct the continuous function basis for the second approach reducing the computational efficiency.
Based on the second idea, a globally
conservative approach has been used in \cite{kaltenbacher10:2}, where
the acoustic sources within the finite element (FE) formulation are first computed on
the fine flow grid. These so-called loads are then
interpolated by a conservative scheme to the acoustic grid. This
approach works with high accuracy in cases where the flow grid is much
finer than the acoustic grid. However, in cases where the flow grid gets
coarser than the acoustic grid, an approach like this fails. To overcome
such problems, we have applied a cell volume weight interpolation and
successfully applied it to the aeroacoustic computation of an axial fan
\cite{kaltenbacher2017computational}. Besides acoustic finite element simulations, similar investigations have been
performed in \cite{Schroeder2016}, where for both the flow and the
acoustic field, a finite volume (FV) scheme has been used. 
To accelerate the calculation for volume sources within a boundary
element (BE) method solving Lighthill's inhomogenous wave equation,
the authors in \cite{croaker2013fast} applied a particle condensation
technique to spatially condense the acoustic source data. 

Based on the first
application of the cut-volume cell weight interpolation \cite{kaltenbacher2017computational}, we
study the acoustic effects of the current approach compared to the limited cell-centroid based 
approach \cite{kaltenbacher10:2}. 
The computational algorithms are outlined using pseudo code that aligns the computational formulas in a step-by-step workflow. Considering an analytic example, we illustrate the benefits of the cut-volume cell weight interpolation for a set of different flow and acoustic discretizations. It clarifies when an application of the cut-volume cell weight interpolation is more beneficial, taking the computational performance and the accuracy of the interpolation into account. Then we apply both methods in a hybrid workflow to an axial fan. The flow simulation and the experimental investigations have already been presented in \cite{kaltenbacher2017computational}. Upon previous findings, we exclusively concentrate on the interpolation between the flow and the acoustic mesh trough a mesh study. We show the increase in computational performance through coarsening the acoustic mesh without loosing numerical accuracy and discuss the suitability of the applied interpolation methods.

The paper is organized as follows: In Sec.
\ref{Sec:Formulation}, we provide the physical basis of the perturbed convective wave equation, and thoroughly describe the formulation of the cut-volume cell approach. 
In Sec. \ref{Application}, we study the effects of two different conservative interpolation schemes on the acoustic results with respect to the acoustic discretization. Finally, in Sec. \ref{sec:conclusion} we conclude
the findings and give additional remarks for future research.
%

\section{Formulation} \label{Sec:Formulation}
Hardin and Pope~\cite{HardinandPope:94} introduced the acoustic/viscous splitting technique for the prediction of flow-induced sound. Afterwards, many scientists applied this idea and derived linear and non linear wave equations~\cite{Shen1999,Ewert:03,Seo2005,Munz2007}. The essence of all these methods is that the flow field quantities are split into compressible and incompressible parts

\begin{eqnarray}
p &=& \bar p + p^{\textrm ic} + p^{\textrm c} = \bar p + p^{\textrm ic} + p^{\textrm a} \\
\bm v &=& \bar{\bm v} + \bm v^{\textrm ic} + \bm v^{\textrm c} =  \bar{\bm v}
+ \bm v^{\textrm ic} + \bm v^{\textrm a}\\
\rho &=& \bar \rho + \rho_1 + \rho^{\textrm a}\,. \label{eq:densityCorr}
\end{eqnarray}
In this sense, the field variables are decomposed into mean ($\bar p$, $\bar{\bm
  v}$, $\bar \rho$) and fluctuating parts. Additionally, the fluctuating parts are further
split into acoustic ($p^{\textrm a}$, $\bm v^{\textrm a}$, $\rho^{\textrm a}$) and flow
components ($p^{\textrm ic}$, $\bm v^{\textrm ic}$). Finally, a density
correction $\rho_1$ is built according to Eq.
\eqref{eq:densityCorr}. 
Introducing an ALE (Arbitrary Lagrangian-Eulerian) description
for the operators 
\begin{equation}
\label{eq:pressRotAPE2}
\ \ \frac{D}{Dt} = \frac{\partial }{\partial t} +  \big(  \overline{ \bm v}
- \bm v_{\textrm r}\big)\cdot \nabla \,,
\end{equation}
where $\bm v_{\textrm r}$ is the relative velocity of the grid, we arrive at the perturbed convective wave equation (PCWE) for rotating systems (see \cite{kaltenbacher2017computational})
\begin{equation}
\label{eq:PCWE}
\frac{1}{c^2} \, \, \frac{D^2\psi^{\textrm a}}{D t^2} - \Delta \psi^{\textrm a} =
- \frac{1}{\bar \rho c^2}\, \frac{D p^{\textrm ic}}{D t}\,.
\end{equation}
This scalar convective wave equation is computationally efficient and describes aeroacoustic sources
generated by incompressible flow structures and wave propagation
through moving media. This formulation reduces the number of unknowns (acoustic pressure
$p^{\textrm a} = \bar \rho \frac{D\psi^{\textrm a}}{D t}$ and particle velocity $\bm v^{\textrm a}= -\nabla \psi^{\textrm a}$) to just a scalar unknown, the acoustic velocity potential $\psi^{\textrm a}$. Based on the PCWE, we verify the source term interpolation of aeroacoustic sources for non-moving and moving meshes.

The derivation of the weak formulation for the finite element method involves a multiplication with a test function $\psi \in H^1(\Omega)$ and an integration over the computational domain $\Omega$. The aeroacoustic source $ f^{{\textrm a}} := - \frac{1}{\bar \rho c^2}\, \frac{D p^{\textrm ic}}{D t}$ in the weak formulation yields
\begin{equation}
\int_{\Omega} \psi f^{{\textrm a}} \mathrm{d}\xi \, .
\label{eq:RHS}
\end{equation}
In order to compute the aeroacoustic sources on the acoustic grid in a hybrid workflow, the integral must be approximated by the field quantities on the flow grid. This naturally leads to an interpolation that conserves the integral property of the aeroacoustic sources used by the finite element simulation.

\subsection{Source term interpolation}\label{Sec:Interpol}
\begin{figure}[hbt]
 \centering
   \subfigure[Cell-centroid based approach for conservative interpolation.]{
     \label{fig:conservInterp1}
     \includegraphics[width=0.65\textwidth,trim={0.1cm 0.1cm 0.1cm 0.1cm},clip]{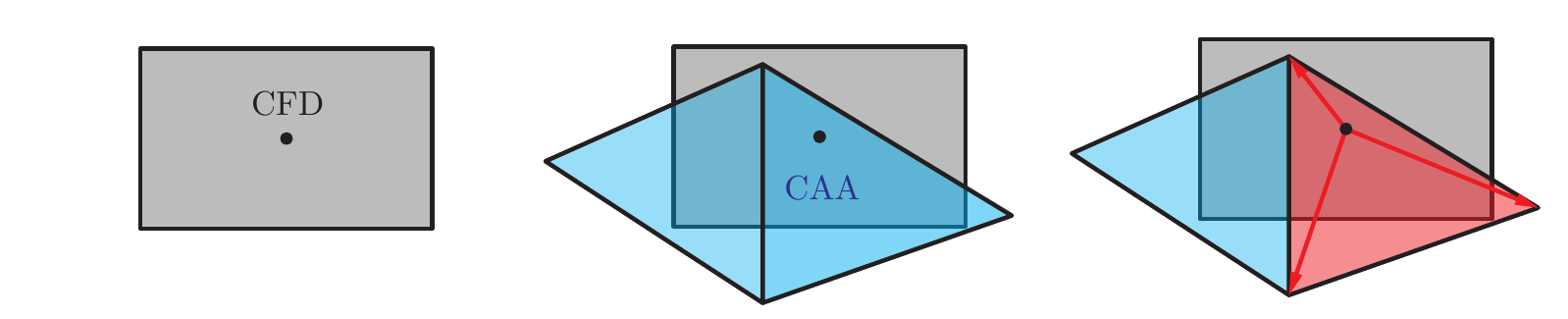}
   } \hspace*{5mm}   
   \subfigure[CFD grid (red) and CAA grid (blue).]{
     \label{fig:InterpMesh}
     \includegraphics[width=0.28\textwidth]{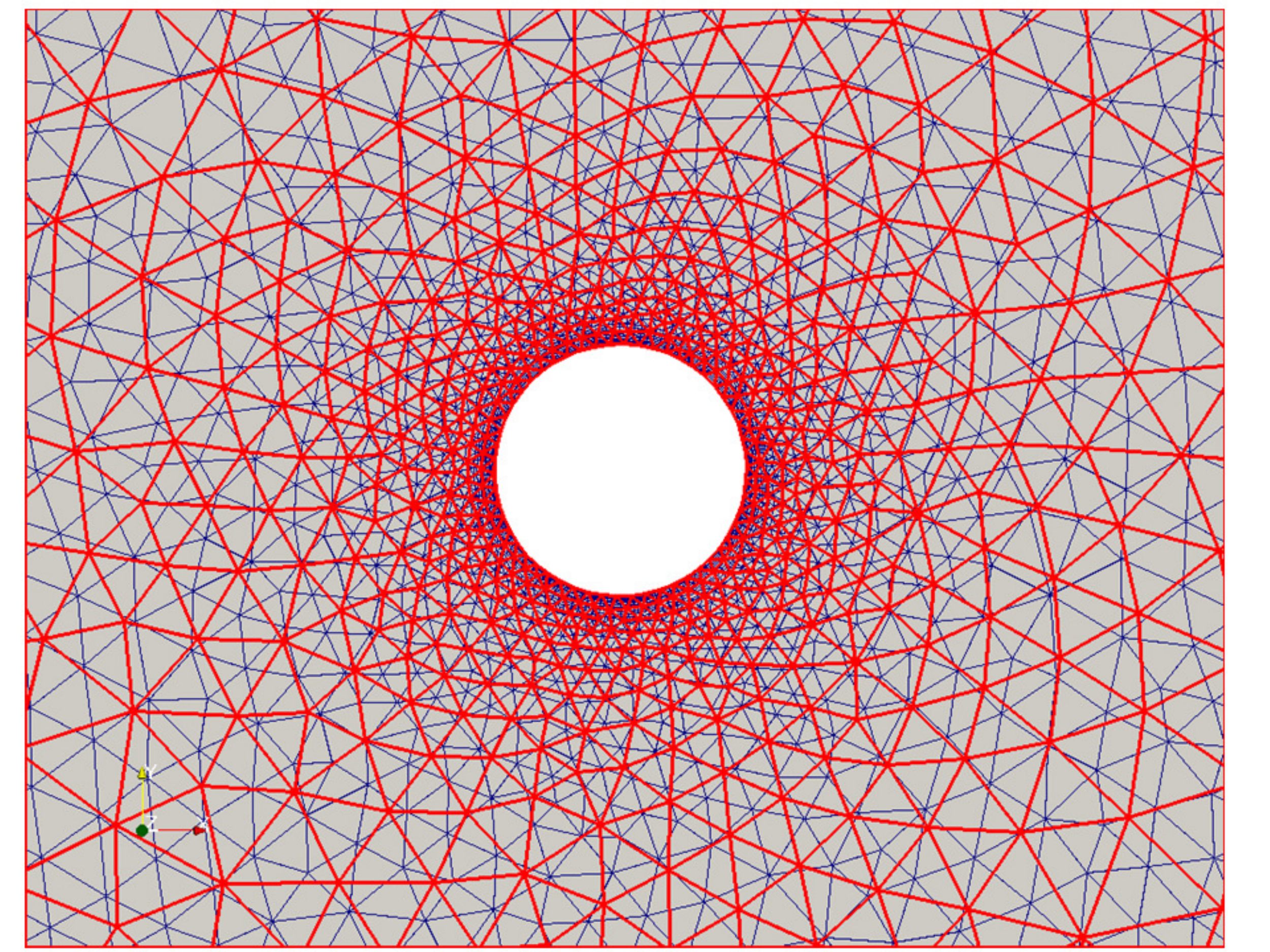}
   }   
    \caption {Cell-centroid based approach for conservative interpolation and
      general mesh sizes of flow and acoustic grid.}
 \label{fig:conservInterpGen}
\end{figure}

Almost every accurate hybrid aeroacoustic approach using finite element method for the acoustic simulation requires a conservative
transformation of the acoustic sources from the 
flow grid (superscript f) to the acoustic grid (superscript a). Starting from \eqref{eq:RHS}, the cell-centroid based approach conserves the energy globally
but approximates the local energy conservation of the finite element right hand side by
\begin{equation}
F_i^{\textrm a} = \int_{E^{\textrm a}} N_i^{\textrm a} (\bm \xi) f^{{\textrm a}} \mathrm{d}\xi \approx \sum_{k \in M^\mathrm{f}} N^{\textrm a}_i (\bm \xi_{E_k^\mathrm{f}}) F_k^{\textrm f}   \, ,
\label{eq:FEMInterpol}
\end{equation}
where $E_k^\mathrm{f}$ denotes the cell of the flow grid, $\bm \xi_{E_k^\mathrm{f}}$ the local coordinate, $M^\mathrm{f}$ the number of flow cells, and the subscript $i$ the node number on the CAA grid.
In order to preserve the acoustic energy, the loads $F_k^{\textrm f}$ of the fine flow grid are interpolated to
the coarser acoustic grid (see Fig. \ref{fig:conservInterp1}).

For an acoustic element, the cell-centroid based approach (Algo.~\ref{alg:centroid}) includes all loads $F_k^{\textrm f}$ that are inside the acoustic finite element ${E^{\textrm a}}$.  This geometrical connection from load location to the acoustic element is accounted for by the set $M^\mathrm{f}$. Based on the geometrical properties, the loads are weighted by the finite element basis functions and summed up onto the acoustic grid. The loads
\begin{equation}
F_k^{\textrm f} =  \bigwedge_{l \rightarrow k} \int_{E_l^\mathrm{f}}  N_l^{\textrm f} (\bm \xi) f^{\textrm f}(\bm \xi) \mathrm{d}\xi
\label{eq:Source}
\end{equation}
are assembled on the CFD mesh in terms of the finite element method, by the assembly operator $\bigwedge$. To reduce the computational complexity of \eqref{eq:Source} and to construct a computationally efficient algorithm, we simplify the integration over the source volume
to a first order volume weighting 
\begin{equation}
F_k^{\textrm f} =  V_{E_k^\mathrm{f}} f_k^{\textrm f}  (\bm \xi_{E_k^\mathrm{f}})
\label{eq:simplifiedSource}
\end{equation}%
and avoid the finite element assembly operator. 
The error introduced by this approximation may be estimated based on a Taylor series and is thus bounded by the product of the characteristic grid size and the gradient of the interpolated quantity, which is assumed to be small for a high quality flow grid.
As illustrated in
Fig. \ref{fig:conservInterp1}, the algorithm has to find the nodal source location $\bm x_k$ inside an acoustic element (set $M^\mathrm{f}$). 
This global position $\bm x_k$ corresponds to a local position
$\bm \xi_{E_k^\mathrm{f}}$ in the reference finite element ($\bm x_k \multimap \bm \xi_{E_k^\mathrm{f}}$). Finally, the loads $F^{\textrm f}_k$ are interpolated to the nodes of the acoustic mesh
using the finite element basis functions $N^{\textrm a}_i$ \cite{kaltenbacher10:2}.

\begin{algorithm}[ht!]
  \caption{Cell-centroid based interpolation}
    \label{alg:centroid}
  \begin{algorithmic}[1]
  \State $\mathbf{F^{\textrm a}} \gets \mathrm{CentroidInterpolation}({E^{\textrm a}})$
    \Statex
    \Function{CentroidInterpolation}{}
      \For{$i$ in ${E^{\textrm a}}$}
      	\State $M^\mathrm{f}=\text{GetAllCFDCellsOfCAAElementByCFDCellCenter}(i)$\quad \Comment{$M^\mathrm{f}$ holds indices of CFD Cells}
      	\For{$k$ in $M^\mathrm{f}$}
      	    \State $V_{E_k^\mathrm{f}}=\text{GetVolumeOfCFDCell(k)}$   	
		    \State $F_k^{\textrm f}=\text{ComputeFEMLoadOnCFD(k)}$ \quad \Comment{Compute \eqref{eq:simplifiedSource}}
		    \State $\boldsymbol{ x}_k=\text{GetCentroidOfCFDCell(k)}$
			\State $F_i^{\textrm a} = \text{InterpolateFEMLoadToCAAMesh}(\boldsymbol{ x}_k)$ \quad \Comment{Finite element interpolation to CAA element \eqref{eq:FEMInterpol}}
		\EndFor		      	
      \EndFor
    \EndFunction
  \end{algorithmic}
\end{algorithm}

This cell-centroid based approach works accurately in cases where the flow grid is much
finer than the acoustic grid. However, the mesh size of flow
computations varies from fine meshes resolving boundary layers to coarse meshes towards regions without flow gradients. Consequently, hybrid aeroacoustics deals with regions where the size of the flow grid is equal to or larger than the acoustic mesh (see Fig. \ref{fig:InterpMesh}). In these cases, the cell-centroid based approach fails locally because it neglects the contributions weighted by the volume.

\begin{figure}[hbt]
    \centering
    \includegraphics[width=0.7\textwidth,trim={0.1cm 0.1cm 0.1cm 0.1cm},clip]{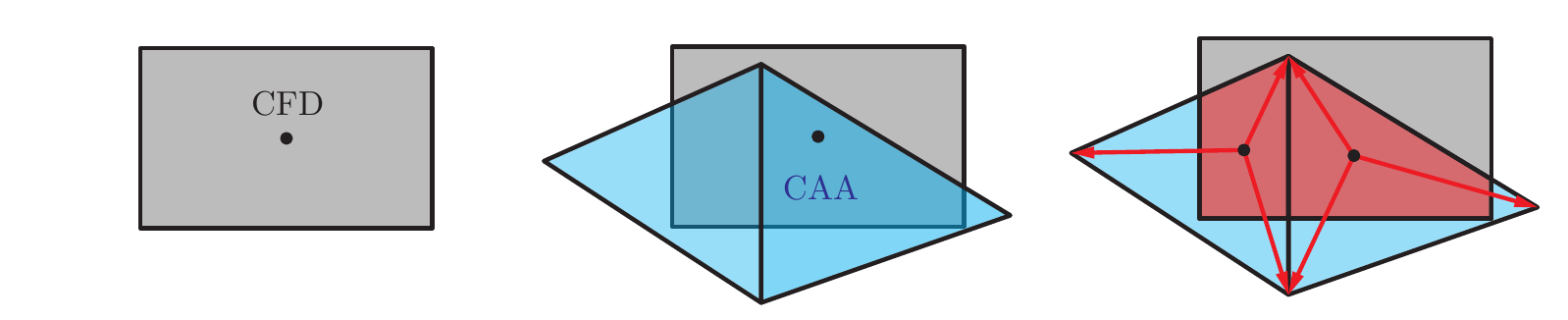}
    \caption{Steps for intersection and interpolation based on cut-volume cell approach.}
    \label{fig:volCut}
\end{figure}

Thus, the improved approach considers for each acoustic volume element $E_l^\mathrm{a}$ the set of flow cells $E^\mathrm{f}$ that intersect \vspace{-0.2cm}
\begin{equation}
E_l^\mathrm{f \cap a} = {E_l^\mathrm{a} \cap E^\mathrm{f}}
\end{equation}
with the respective acoustic volume element. This improved approach conserves the energy globally as well as locally for different mesh sizes and calculates the finite element right hand side by
\begin{equation}
\int_{E^{\textrm a}} N_i^{\textrm a} (\bm \xi) f^{{\textrm a}} \mathrm{d}\xi = \sum_{k \in M^\mathrm{f \cap a}} N^{\textrm a}_i (\bm \xi_k) F_k^\mathrm{f \cap a} \,.
\label{eq:FEMInterpol2}
\end{equation}
Based on this intersection, the loads $F_k^\mathrm{f \cap a} $ are volume-weighted,
\begin{equation}
F_k^\mathrm{f \cap a} =  \bigwedge_{l \rightarrow k} \int_{E_l^\mathrm{f \cap a}}  N_l^{\textrm f}   (\bm \xi) f^{\textrm f}(\bm \xi) \mathrm{d}\xi \, .
\label{eq:Source2}
\end{equation}
To avoid the finite element assembly operator in \eqref{eq:Source2}, we assume that the aeroacoustic source term is constant over the
fluid cell. Therefore, the integral reduces to a multiplication of the intersection volume $V_\mathrm{c} \multimap V_{E_l^\mathrm{f \cap a}}$ with the source density $f_l^{\textrm f}$ at the volumetric centroid $\bm x_\mathrm{c} \multimap \bm \xi_{E_l^\mathrm{f \cap a}}$ of the intersection polyhedron
\begin{equation}
F_k^\mathrm{f \cap a} =  V_{E_k^\mathrm{f \cap a}} f_k^{\textrm f}   (\bm \xi_{E_k^\mathrm{f \cap a}})\,.
\label{eq:simplifiedSource2}
\end{equation}%
The algorithmic workflow is depicted in Fig.~\ref{fig:volCut}. As opposed to the cell-centroid based approach, we determine two intersecting cells; we compute the volumetric center $\bm x_\mathrm{c} \multimap \bm \xi_{E_l^\mathrm{f \cap a}}$ and the volume of the intersection polyhedron $V_\mathrm{c} \multimap V_{E_l^\mathrm{f \cap a}}$.  
This additional intersection information allows us to add the source contribution
weighted with the intersection volume to the nodes of the acoustic
element using the finite element basis functions evaluated at
the center of the polyhedron (see Fig.~\ref{fig:volCut}). Thereby, we cover cases in which acoustic
elements are smaller and embedded into a single flow cell.
\begin{algorithm}[ht!]
  \caption{cut-volume cell interpolation}
    \label{alg:cutcell}
  \begin{algorithmic}[2]
  \State $\mathbf{F^{\textrm a}} \gets \mathrm{CutVolumeCellInterpolation}({E^{\textrm a}})$
    \Statex
    \Function{CutVolumeCellInterpolation}{}
      \For{$i$ in ${E^{\textrm a}}$}
        \State $M^\mathrm{f \cap a}=\text{IntersectCFDGridWithCAACell}(i)$\quad \Comment{$M^\mathrm{f \cap a}$ holds indices of CFD Cells}
      	\For{$k$ in $M^\mathrm{f \cap a}$}
      	    \State $V_{E_l^\mathrm{f \cap a}}=\text{GetIntersectionVolumeOfCFDCell(k)}$   	
		    \State $F_k^\mathrm{f \cap a}=\text{ComputeFEMLoadOnCFD(k)}$ \quad \Comment{Compute \eqref{eq:simplifiedSource2}}
		    \State $\boldsymbol{ x}_k=\text{GetCentroidOfIntersectionCell(k)}$
			\State $F_i^{\textrm a} = \text{InterpolateFEMLoadToCAAMesh}(\boldsymbol{ x}_k)$ \quad \Comment{Finite element interpolation to CAA element \eqref{eq:FEMInterpol2}}
		\EndFor		      	
      \EndFor
    \EndFunction
  \end{algorithmic}
\end{algorithm}

For variable acoustic sources within one fluid cell, the full finite element framework has to be carried out during the conservative interpolation. An additional interpolation or regression step must be executed to construct the continuous function basis. Then the source density $f_l^{\textrm f}$ could be raised from the lowest order polynomial space to an arbitrary order. As a consequence, the integral \eqref{eq:Source2} must be computed which involves matrix multiplications instead of scalar multiplications. Therefore, the centroid position and the volume of a fluid cell have to be transformed to the reference finite element.

\begin{figure}
 \centering
   \subfigure[Cell-centroid, $\Gamma = 0.5$]{
   \includegraphics[scale=0.98]{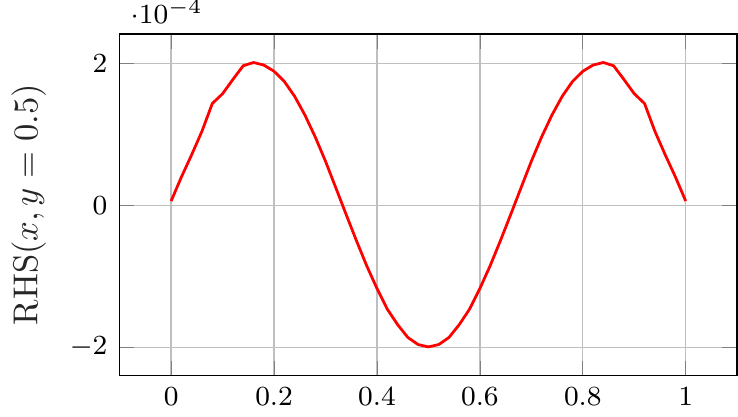}
   } \hspace*{5mm}   
   \subfigure[Cut, $\Gamma = 0.5$]{
   \includegraphics[scale=0.98]{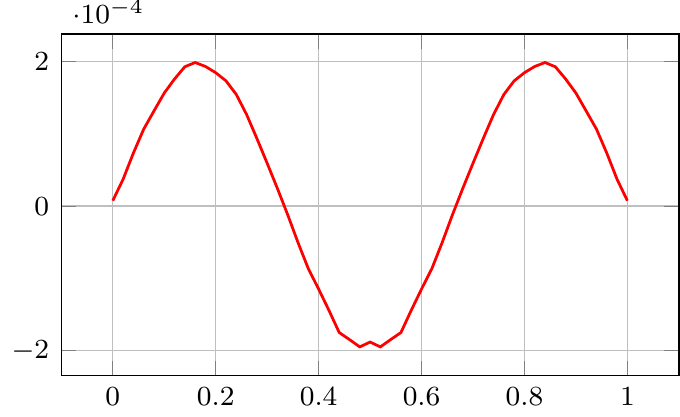}
   }\\
   \subfigure[Cell-centroid, $\Gamma = 1$]{
   \includegraphics[scale=0.98]{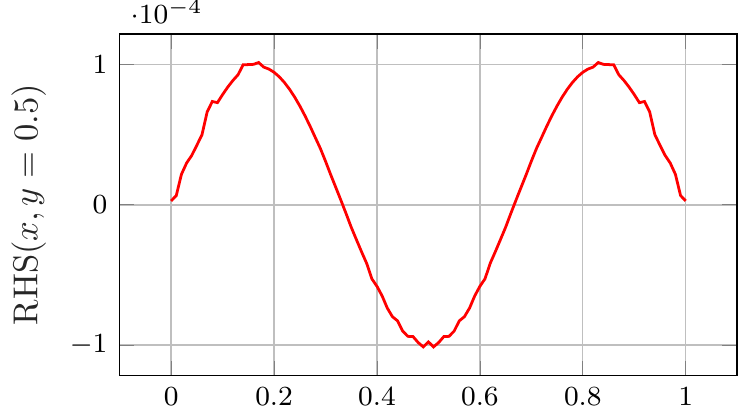}
   } \hspace*{5mm}   
   \subfigure[Cut, $\Gamma = 1$]{
   \includegraphics[scale=0.98]{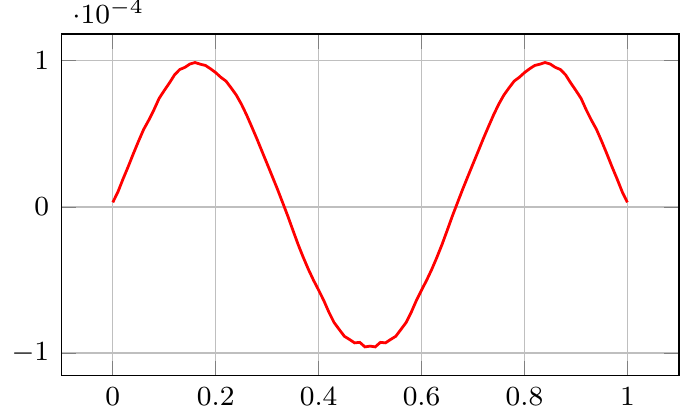}
   }\\
   \subfigure[Cell-centroid, $\Gamma = 2$]{
   \includegraphics[scale=0.98]{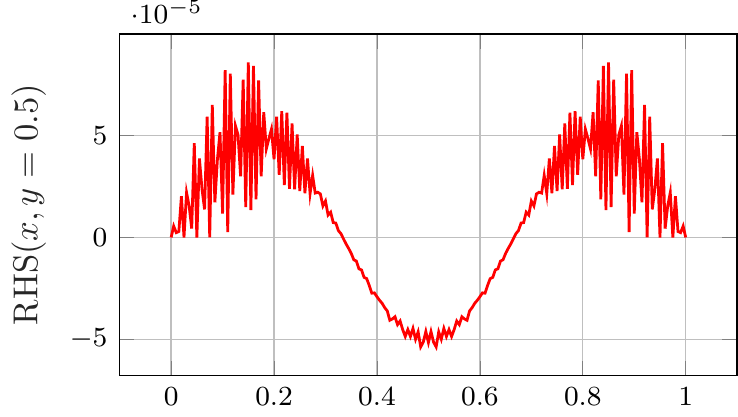}
   } \hspace*{5mm}   
   \subfigure[Cut, $\Gamma = 2$]{
   \includegraphics[scale=0.98]{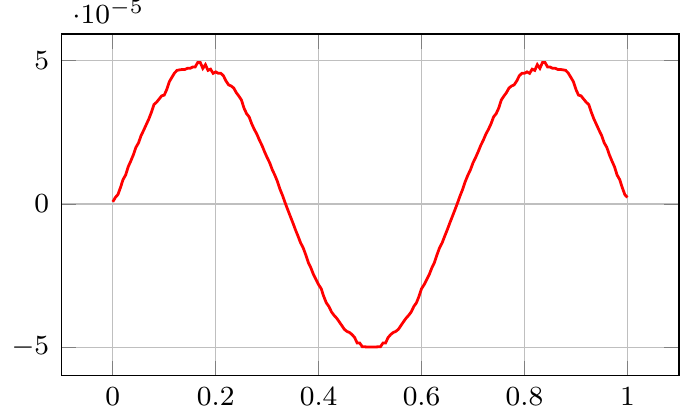}
   }\\
   \subfigure[Cell-centroid, $\Gamma = 30$]{
   \includegraphics[scale=0.98]{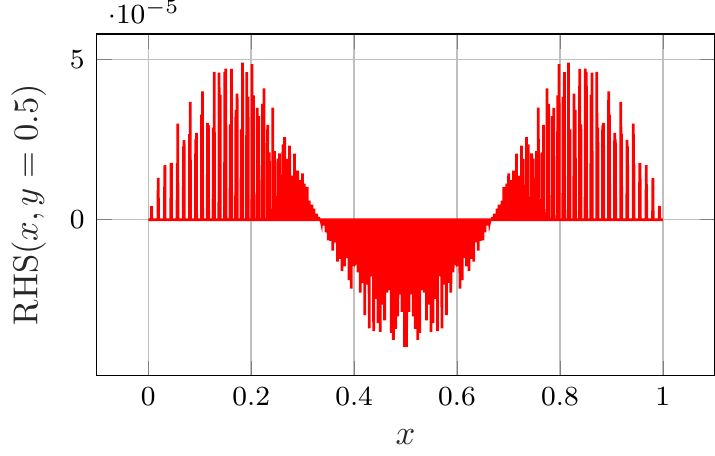}
   } \hspace*{5mm}   
   \subfigure[Cut, $\Gamma = 30$]{
   \includegraphics[scale=0.98]{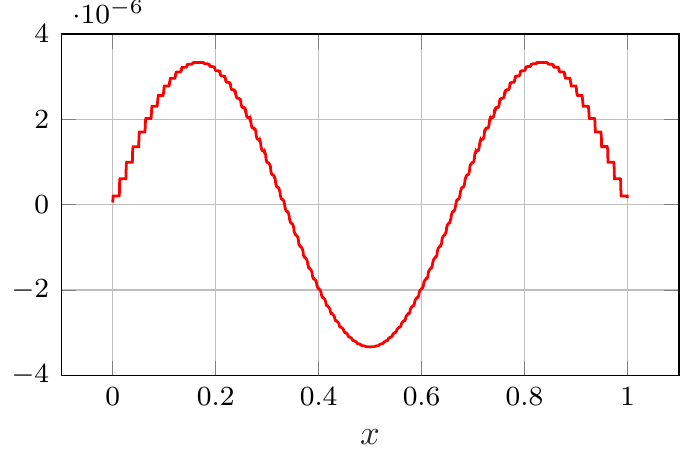}
   }   
    \caption{FE nodal right hand side values at $y=0.5$ with respect to the distance in $x$-direction. The number of nodes of the CFD grid along the evaluation line is 101; whereas the number of nodes of the CAA grid along the line is based on the mesh ratio $\Gamma$. $\Gamma<1$: the CFD grid is finer than the CAA mesh; $\Gamma>1$: the CFD grid is coarser than the CAA mesh. The cell-centroid based interpolation for different mesh ratios is illustrated in (a, c, e, g) and the cut-volume cell approach is illustrated in (b, d, f, h).}
 \label{fig:conservInterpRHS}
\end{figure}

The cell-centroid based conservative interpolation and the cut-volume cell approach are compared and verified against an analytic function 
\begin{equation}
	f\left(\boldsymbol{ x}\right) = \sin\left (3 \pi x\right)
\end{equation}
on the domain $\Omega \in [0,1]^2$. Considering the cell-centroid based approach, energy errors will occur for large mesh ratios $\Gamma$
\begin{equation}
	\Gamma = \frac{N_{\textrm {CAA}}}{N_{\textrm {CFD}}}\, ,
\end{equation}
where $N_{\textrm {CAA}}$ is the number of nodes on the edges of the CAA mesh and $N_{\textrm {CFD}}$ is the number of nodes on the edges of the CFD mesh. Figure \ref{fig:conservInterpRHS} shows how the FE nodal right hand side values develop on a line at $y=0.5$. Along the  evaluation line, the number of nodes of the CFD grid are constant for the different evaluations (101 nodes). In contrast to that, the number of nodes along the CAA mesh vary according to the mesh ratio. If $\Gamma<1$, the CFD grid is finer than the CAA mesh; for $\Gamma>1$, the CFD grid is coarser than the CAA mesh. The cell-centroid based interpolation performs satisfactorily for low mesh ratios where the CFD grid is finer than the CAA mesh. For larger mesh ratios, the cell-centroid based interpolation exhibits spurious modes and energy is transferred to higher wave lengths (unphysical). Overall, the cut-volume cell approach has the desired conservative properties for all mesh ratios.

Based on Parceval's theorem, the relative energy content of the real wavelength $\lambda_{{\textrm a}}$ compared to the total energy is evaluated by the energy ratio $e_{\textrm r}$
\begin{equation}
e_{\textrm r} = \frac{\left(\hat{D}_{\textrm {RHS}}(\lambda_{\textrm a})\right)^2}{\sum_{k=1}^N \left(\hat{D}_{\textrm {RHS}}(\lambda_k)\right)^2} \,.
\end{equation}
$\hat{D}_{\textrm {RHS}}$ are the amplitudes of the discrete wave number transform at the corresponding wavenumber. Table \ref{tab:analyticEnergy} compares the cell-centroid based conservative interpolation (cell-centroid) to the conservative cut-volume cell interpolation (cut-volume cell).
As expected, the more accurate conservative cut-volume cell interpolation significantly outperforms the cell-centroid based procedure. However, the computational demand increases with the number of cell intersections. Compared to the setup time of the cell-centroid based approach, the setup time for the cut-volume cell intersection takes about 50 times as long but overall, the computational demand increases linearly.

Additionally to the mesh ratio $\Gamma$, Tab.~\ref{tab:analyticEnergy} lists the coverage ratio $\varepsilon$, which describes the ratio of CAA cells that are covered by CFD cells relative to the total number of CAA cells. For example, the CAA mesh in Fig. \ref{fig:conservInterp1} has a coverage ratio of $\varepsilon=0.5$; reddish cells are covered and bluish cells are not covered. It is obvious that the energy ratio $e_{\textrm r}$ for the cell-centroid based interpolation decreases with decreasing coverage ratio $\varepsilon$. Closer examination of the correlation between $e_{\textrm r}$ for the cell-centroid based interpolation and $\varepsilon$ revealed that $e_{\textrm r} \sim 1-(1-\varepsilon)^3$.

\begin{table}[ht!]
\centering
\caption{\label{tab:analyticEnergy}Ratio of the energy corresponding to the actual wavelength and the total energy for the cell-centroid based procedure and the cut volume-cell procedure, respectively.}
\begin{tabular}{rrrrrr}
\toprule
 &  &   &  & CPU time & CPU time\\
$\Gamma$ & Coverage ratio $\varepsilon$ & Cell-centroid $e_{\textrm r}$  & Cut volume-cell $e_{\textrm r}$ & cell-centroid & cut volume-cell\\
\midrule
0.5 & 99.98 \% & 99.28 \% & 99.22 \% & \SI{0.282}{\second} & \SI{4.387}{\second}\\
1 & 87.18 \% & 99.17 \% & 98.92 \% & \SI{0.279}{\second} & \SI{6.61}{\second}\\
2 & 50.29 \% & 85.48 \% & 98.79 \% & \SI{0.368}{\second} & \SI{10.983}{\second}\\
5 & 20.2 \% & 52.55 \% & 98.85 \% & \SI{0.562}{\second} & \SI{23.938}{\second}\\
10 & 10.1 \% & 31.72 \% & 98.94 \% & \SI{0.835}{\second} & \SI{46.158}{\second}\\
20 & 5.05 \% & 15.73 \% & 99.04 \% & \SI{1.457}{\second} & \SI{88.403}{\second}\\
30 & 3.37 \% & 11.03 \% & 99.1 \% & \SI{2.048}{\second} & \SI{129.853}{\second}\\
\bottomrule  
\end{tabular}
\end{table}

\section{Application} \label{Application}
\begin{figure}[hbt]
 \centering
   \subfigure[Geometry of the investigated fan.]{
     \label{fig:rotor}
     \includegraphics[width=0.35\textwidth]{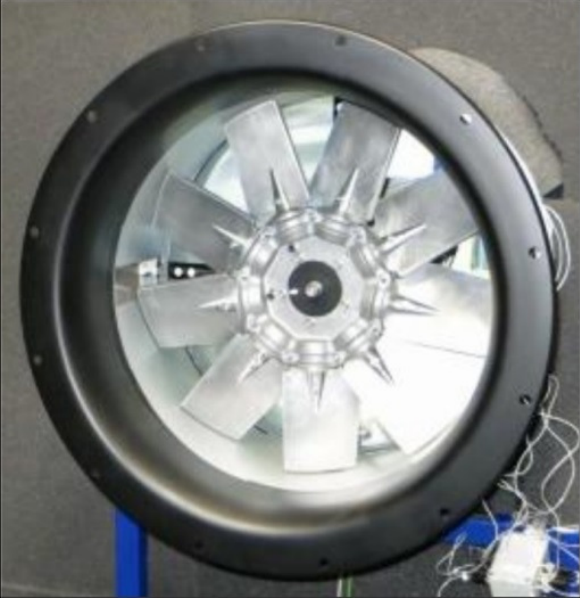}
   } \hspace*{5mm}   
   \subfigure[Geometry of acoustic domain with the rotating source region in yellow.]{
     \label{fig:caaDomain}
     \includegraphics[width=0.55\textwidth]{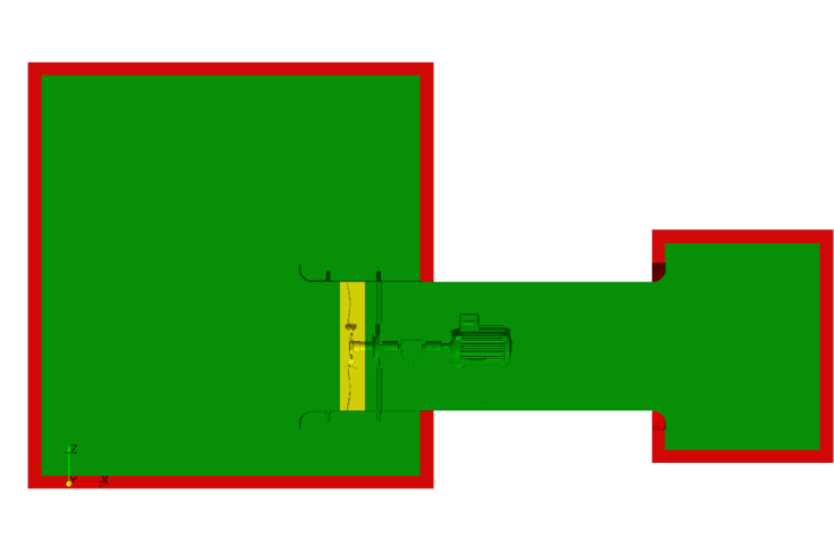}
   }   
    \caption{Investigated fan.}
 \label{fig:conservInterp}
\end{figure}

The aforementioned methods are now applied to a fan (see Fig. \ref{fig:rotor}) that was already presented in \cite{kaltenbacher2017computational}. It has a diameter of 0.5\,m, consists of 9 flat plates, is operated at 1500\,rpm and a volume flow of 1.3\,m$^3/$s, and is mounted in a duct together with the motor.
The incompressible pressure required to compute the source terms according to Eq.~(\ref{eq:PCWE}) was obtained by a detached eddy CFD simulation. The source terms are computed on the CFD mesh (33\,M cells, with 22\,M cells in the rotating region and a maximum cell size of 1\,mm) and then interpolated to the acoustic domain, which is displayed in Fig. \ref{fig:caaDomain}. The yellow region is the rotating region including the fan, where the main acoustic source terms occur. The green domain is a stationary propagation region and the red region is a perfectly matched layer to account for the free radiation condition. The reflecting surfaces of the nozzle, duct, strut, driving shaft, and motor are modeled as acoustically hard walls. After the interpolation, the acoustic propagation computation was done with our in-house FE solver CFS++ \cite{kaltenbacher2015numerical}.

\subsection{Application of the source term interpolation} \label{sec:AP3000interpolation}
\begin{figure}[hbt!]
    \centering
    \includegraphics[width=0.8\linewidth]{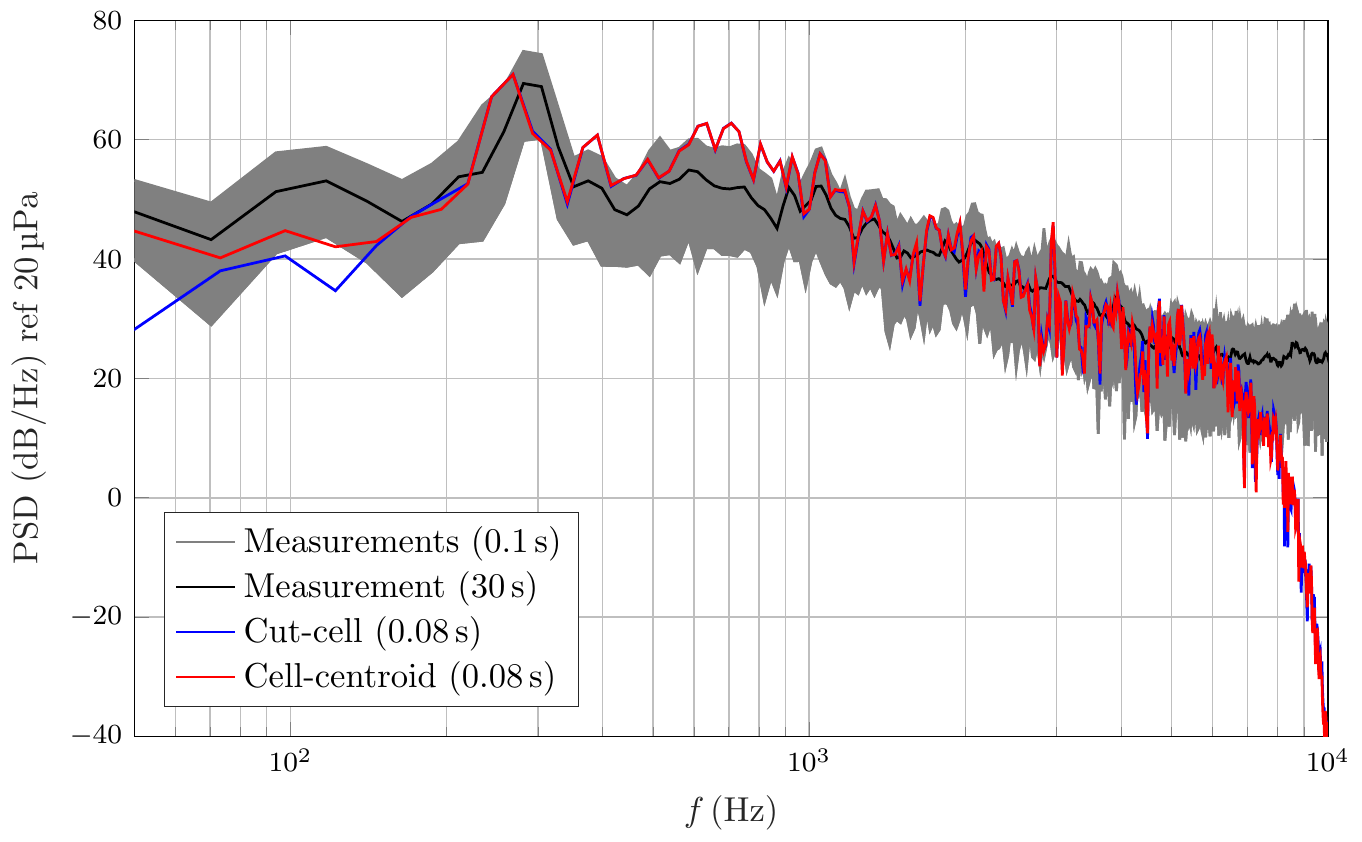}
    \caption{Comparison of the PSD using the cell-centroid based and the cut-volume cell interpolation algorithm with measurement results.}
    \label{fig:comparisonStandardAndCut}
\end{figure}

To investigate the behavior of the interpolation algorithms, the acoustic source term was computed using the incompressible pressure and its gradient, which were both obtained from the CFD simulation. 
The source term was interpolated to the acoustic mesh using the two different interpolation approaches. The acoustic propagation simulation was done with the same solver settings as presented in \cite{kaltenbacher2017computational}. The results of the two simulations are compared with measurement results in Fig. \ref{fig:comparisonStandardAndCut}. To make the measurement and the simulation comparable, the power spectral density (PSD) was used. All depicted results show the PSD at a position of 1\,m in front of the inlet nozzle.
The black line represents a measurement with a measurement time of 30\,s. The gray lines are measurement results with a measurement time of 0.1\,s, which is close to the simulation time of 0.08\,s. Both simulation results have a steep drop above 6\,kHz where the mesh resolution in the propagation domain becomes coarse compared to the acoustic wave length. The red curve shows the result of the cell-centroid based approach. It predicts the PSD well over the whole frequency range and therefore is useable in cases with complex geometry and distorted grid cells of varying cell sizes. This statement is true as long as the CFD cells are smaller than the CAA cells. The blue curve represents the result of the cut-volume cell approach, which overall show good agreement with the measurement. This application demonstrates that the cut volume-cell based interpolation can deal with complex geometries arising in real applications. 
The intersection algorithm has an additional computation time of about 45 minutes and heavily depends on the number of CFD cells. Table \ref{tab:meshes} shows a small increase of the intersection time with increasing CAA elements. Based on the coverage ratio, the computations can be automated to switch from the ordinary cell-centroid based interpolation to the more advanced interpolation if necessary.

\begin{figure}[hbt]
 \centering
   \subfigure[Different meshes used for the computation.]{
     \label{fig:comparisonMeshMeshes}
     \includegraphics[width=0.43\textwidth]{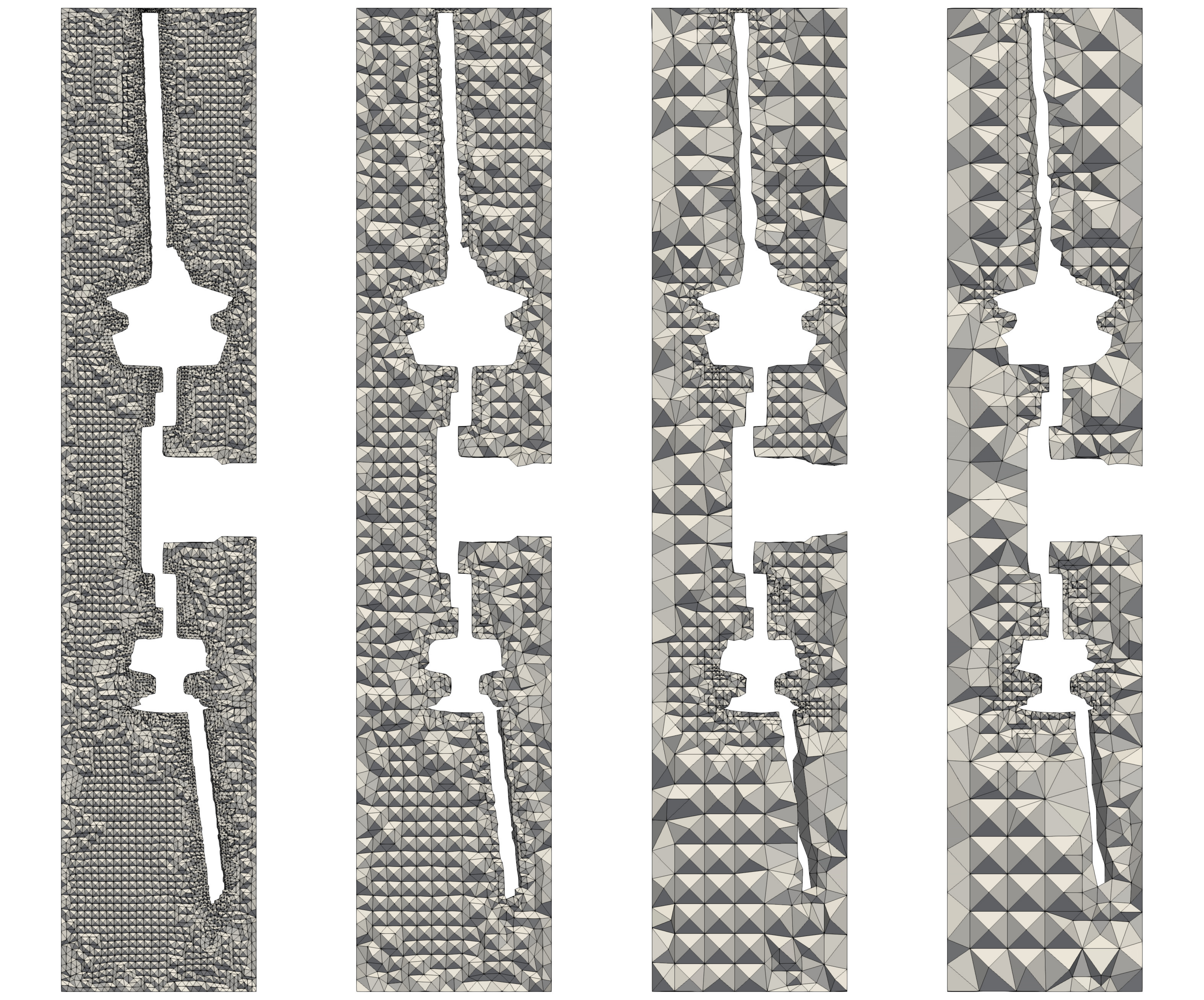}
   } \hspace*{5mm}   
   \subfigure[Acoustic power spectral density for different discretizations.]{
     \label{fig:comparisonMeshResult}
     \includegraphics[width=0.50\textwidth]{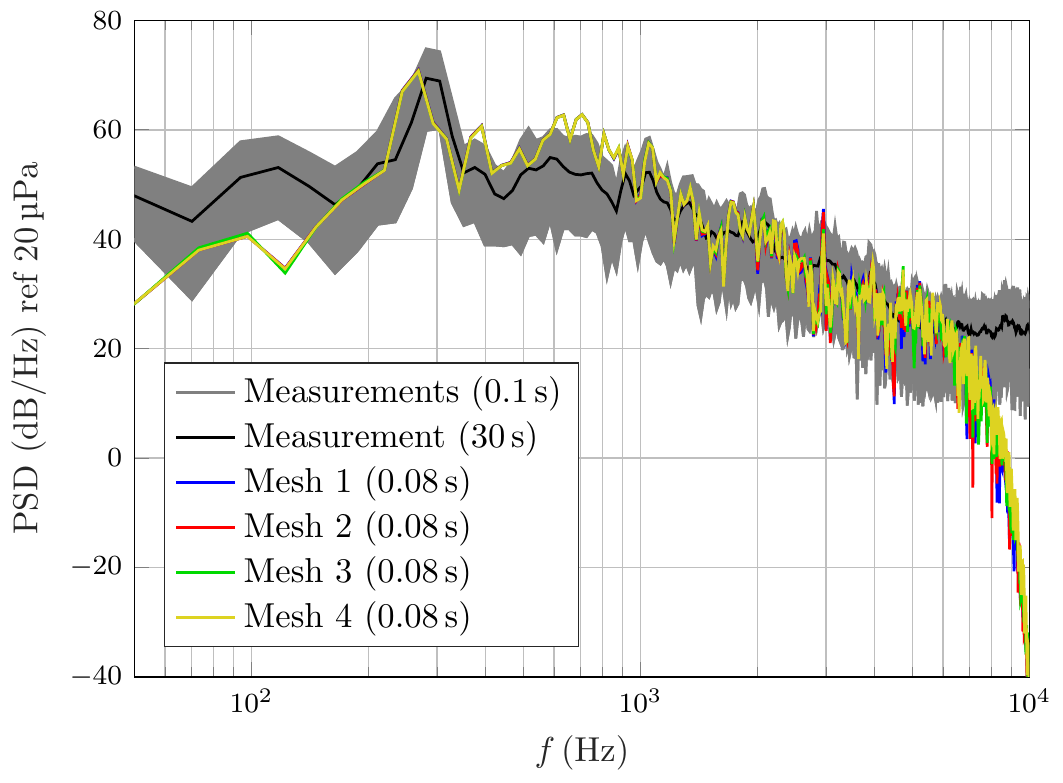}
   }   
    \caption{Comparison of different discretizations (from the finest mesh 1 to the coarsest mesh 4) using the cut-volume cell based interpolation.}
 \label{fig:comparisonMesh}
\end{figure}

The same setup as before was used to investigate the influence of the CAA grid resolution on the cut-volume cell algorithm. This time, the cut-volume cell algorithm was used to interpolate the acoustic source on four different grids in the rotating region. All grids are tetrahedral and have different grid sizes. The numbers of elements in the source region are shown in Tab. \ref{tab:meshes}, where mesh 1 is the finest and mesh 4 is the coarsest mesh. The mesh of the propagation domain was not changed for the different simulations and counted 1804377 nodes.
\begin{table}[ht!]
\centering
\caption{\label{tab:meshes}Different meshes used in the source region to investigate the cut-volume cell interpolation and the cell-centroid based interpolation. The coverage ratio measures the relative coverage of CAA cells by CFD cells for the cell-centroid based interpolation. The intersection time describes the excess CPU time of the cut-volume cell interpolation, due to the intersection algorithm.}
\begin{tabular}{rrrrrr}
\toprule
& Source region & Source region & \\
& elements & nodes & Max. element size & Coverage ratio $\varepsilon$ & Intersection time \\
\midrule
Mesh 1  & 7690908    & 1322937   & 7\,mm  & 85.24 \%  & 49\,min \\
Mesh 2  & 1021697    & 181186    & 12\,mm & 99.78 \%  & 45\,min \\
Mesh 3  & 555562     & 96535     & 18\,mm & 92.87 \%  & 42\,min \\
Mesh 4  & 337013     & 58708     & 24\,mm & 95.14 \%  & 41\,min \\
\bottomrule  
\end{tabular}
\end{table}
The meshes of the rotating region are displayed in Fig. \ref{fig:comparisonMeshMeshes}, where the meshes become coarser from left to right. The minimum cell size of the CAA mesh occurs in the tip gap, where the size is 1\,mm for all CAA meshes. This is necessary to resolve the gap with at least two elements, independently of the discretization of the rest of the mesh. Therefore, the elements in the tip gap dominate the total number of elements for coarse meshes and therefore limit the minimal number of elements. The maximum CFD cell size in this location is 0.2\,mm, which yields a ratio of CAA/CFD $= 5$.

The PSD that resulted from the CAA computation using the interpolated source term and the PSD of the measurements are displayed in Fig. \ref{fig:comparisonMeshResult}. 
The dominant peak at a frequency of $\unit[300]{Hz}$ is a result of both the interaction of the tip gap flow and the blade passing frequency \cite{Kroemer2017A}. It should be emphasized that these frequencies are not the same; however, due to the frequency resolution used in this case they merge to one peak. This peak is well represented in both frequency and amplitude. The colored lines are the results of the numerical simulations with the different meshes. In the low frequency range, the simulations underestimate the PSD which might be due to the very short numerical simulation time. The trend in the higher frequency range is in good agreement with the measurement. Above a frequency of 6\,kHz, the PSD drops since the mesh density in the propagation region is too coarse to resolve the respective wave lengths and the time-stepping scheme with controlled dispersion (Hilber–Hughes–Taylor) numerically 
damps acoustic waves of higher frequency.
All simulations deliver almost identical results up to a frequency of 1\,kHz. Above this frequency it can be seen that the finer meshes lead to more distinctive peaks but the deviation is rather small, even for the coarsest mesh. The deviation in the high frequency range is assumed to have its origin in the mesh density of the source region, where the coarsest mesh meets its limitations to resolve the geometric shape of the source terms.
These results verify the robustness of the cut-volume cell interpolation algorithm.
Furthermore, the reduction of the nodes in the rotation region decreases the overall number of nodes. This reduces the size of the FE system and therefore the computational time. For the finest mesh, the total CPU time was 930\,h and for the coarsest the total CPU time reduced to 296\,h. The strong reduction in CPU time is due to the fact that the finite element simulation time is proportional to the square of the number of nodes.


\section{Conclusions} \label{sec:conclusion}
In this paper, a method for the conservative interpolation of source terms was shown for the application in a hybrid aeroacoustic framework. The interpolation method shown improves the cell-centroid based interpolation method by a cut-volume cell approach. This approach takes the geometry of the interpolation cell into account and therefore improves robustness for meshes with skewed cells and different cell sizes, especially when the CAA grid becomes smaller than the CFD grid. 
Comparing the two interpolation methods, we show how energy is conserved or transferred from the interpolated source mode to artificial modes. 
Therefore, a quality measure of the energy conservation is developed by the coverage ratio. 
This coverage ratio computes the ratio of CAA cell that are covered by a CFD cell and the total number of CAA cells for the cell-centroid interpolation. 
Based on this simple mesh metric, we can judge the quality of a simple source term calculation procedure and switch to a more sophisticated algorithm if necessary.

The application of the cut-volume cell interpolation method was shown on an axial fan. The acoustic results were in good agreement with measurements. Also, the cell-centroid based interpolation method predicts the PSD well. 
But for the finest mesh, the coverage ratio decreases which indicates a possible error source for the cell-centroid based interpolation.
The robustness of the interpolation method was shown for different mesh ratios with a maximum error $e_{\textrm r} < 2\%$. 
This was also shown in the application to the axial fan where the acoustic result changed very slightly even for a reduction of elements in the source region of more than 20 times. 
Overall, the number of unknowns dropped by a factor of 1.67 and the computational time of the CAA simulation was reduced by a factor of 3.14. 
In the future, we plan to investigate an upper bound for the CAA cell size without changing the radiation characteristics of typical noise sources.


\begin{thebibliography}{15}
\newcommand{\enquote}[1]{``#1''}

\bibitem[{Crighton(1993)}]{Crighton:92}
Crighton, D., \enquote{Computational aeroacoustics for low Mach number flows,}
  \emph{Computational aeroacoustics}, Springer, 1993, pp. 50--68.

\bibitem[{Wagner et~al.(2007)Wagner, H\"uttl, and Sagaut}]{Wagner2007}
Wagner, C., H\"uttl, T., and Sagaut, P. (eds.), \emph{Large-Eddy Simulation for
  Acoustics}, Cambridge University Press, 2007.

\bibitem[{Caro et~al.(2009)Caro, Detandt, Manera, Mendonca, and
  Toppinga}]{Caro2009}
Caro, S., Detandt, Y., Manera, J., Mendonca, F., and Toppinga, R.,
  \enquote{{Validation of a New Hybrid CAA Strategy and Application to the
  Noise Generated by a Flap in a Simplified HVAC Duc},} \emph{15th AIAA/CEAS
  Aeroacoustics Conference}, 2009.

\bibitem[{Schr\"oder et~al.(2016)Schr\"oder, Silkeit, and
  Estorff}]{Schroeder2016}
Schr\"oder, T., Silkeit, P., and Estorff, O., \enquote{Influence of source term
  interpolation on hybrid computational aeroacoustics in finite volumes,}
  \emph{InterNoise 2016}, 2016, pp. 1598--1608.
  
  \bibitem[{Piellard and Bailly(2009)}]{Piellard09}
Piellard, M., and Bailly, C., \enquote{A hybrid method for Computational
  Aeroacoustic applied to internal flows,} \emph{NAG-DAGA International
  Conference on Acoustics}, 2009.

\bibitem{hardy1971multiquadric}
R.~L. Hardy, ``Multiquadric equations of topography and other irregular
  surfaces,'' {\em Journal of geophysical research}, vol.~76, no.~8,
  pp.~1905--1915, 1971.

\bibitem{hardy1990theory}
R.~L. Hardy, ``Theory and applications of the multiquadric-biharmonic method 20
  years of discovery 1968--1988,'' {\em Computers \& Mathematics with
  Applications}, vol.~19, no.~8-9, pp.~163--208, 1990.

\bibitem{rendall2008unified}
T.~Rendall and C.~Allen, ``Unified fluid--structure interpolation and mesh
  motion using radial basis functions,'' {\em International Journal for
  Numerical Methods in Engineering}, vol.~74, no.~10, pp.~1519--1559, 2008.

\bibitem{cordero2014radial}
M.~Cordero-Gracia, M.~G{\'o}mez, and E.~Valero, ``A radial basis function
  algorithm for simplified fluid-structure data transfer,'' {\em International
  Journal for Numerical Methods in Engineering}, vol.~99, no.~12, pp.~888--905,
  2014.

\bibitem[{Kaltenbacher et~al.(2010)Kaltenbacher, Escobar, Ali, and
  Becker}]{kaltenbacher10:2}
Kaltenbacher, M., Escobar, M., Ali, I., and Becker, S., \enquote{{Numerical
  Simulation of Flow-Induced Noise Using LES/SAS and Lighthill's Acoustics
  Analogy},} \emph{International Journal for Numerical Methods in Fluids},
  Vol.~63, No.~9, 2010, pp. 1103--1122.

  
  \bibitem[{Kaltenbacher, et~al.(2017)Kaltenbacher, H{\"u}ppe, Reppenhagen, Zenger, and
  Becker}]{kaltenbacher2017computational}
Kaltenbacher, M. and H{\"u}ppe, A.s and Reppenhagen, A. and Zenger, F. and Becker, S., \enquote{Computational aeroacoustics for rotating systems with application to an axial fan,}
  \emph{AIAA journal}, Vol.~55, No.~11, 2017, pp. 3831--3838.
  
\bibitem[{Croaker et~al.(2013)Croaker, Kessissoglou, Kinns, and
  Marburg}]{croaker2013fast}
Croaker, P., Kessissoglou, N., Kinns, R., and Marburg, S., \enquote{Fast
  Low-Storage Method for Evaluating Lighthill’s Volume Quadrupoles,}
  \emph{AIAA journal}, Vol.~51, No.~4, 2013, pp. 867--884.

\bibitem[{Hardin and Pope(1994)}]{HardinandPope:94}
Hardin, J., and Pope, D., \enquote{An acoustic/viscous splitting technique for
  computational aeroacoustics,} \emph{Theoretical and Computational Fluid
  Dynamics}, Vol.~6, No. 5-6, 1994, pp. 323--340.

\bibitem[{Shen and S{\o}rensen(1999)}]{Shen1999}
Shen, W.~Z., and S{\o}rensen, J.~N., \enquote{Aeroacoustic modelling of
  low-speed flows,} \emph{Theoretical and Computational Fluid Dynamics},
  Vol.~13, No.~4, 1999, pp. 271--289.

\bibitem[{Ewert and Schr{\"o}der(2003)}]{Ewert:03}
Ewert, R., and Schr{\"o}der, W., \enquote{Acoustic perturbation equations based
  on flow decomposition via source filtering,} \emph{Journal of Computational
  Physics}, Vol. 188, No.~2, 2003, pp. 365--398.

\bibitem[{Seo and Moon(2005)}]{Seo2005}
Seo, J., and Moon, Y.~J., \enquote{{Perturbed compressible equations for
  aeroacoustic noise prediction at low mach numbers},} \emph{AIAA journal},
  Vol.~43, No.~8, 2005, pp. 1716--1724.

\bibitem[{Munz and Roller(2007)}]{Munz2007}
Munz, M., C.-D.and~Dumbser, and Roller, S., \enquote{{Linearized acoustic
  perturbation equations for low Mach number flow with variable density and
  temperature},} \emph{Journal of Computational Physics}, Vol. 224, No.~1,
  2007, pp. 352--364.
  
\bibitem{kaltenbacher2015numerical}
M.~Kaltenbacher, {\em {Numerical Simulation of Mechatronic Sensors and
  Actuators: Finite Elements for Computational Multiphysics}}.
\newblock Springer Berlin Heidelberg, 2015.

\bibitem[{Kr{\"o}mer et~al.(2018)Kr{\"o}mer, M{\"u}ller, and
  Becker}]{Kroemer2017A}
Kr{\"o}mer, F., M{\"u}ller, J., and Becker, S., \enquote{Investigation of
  aeroacoustic properties of low-pressure axial fans with different blade
  stacking,} \emph{AIAA Journal}, Vol.~56, No.~4, 2018, pp. 1507--1518.

\end{thebibliography}
\end{document}